\documentstyle[11pt,twoside]{article}
\newlength{\dinwidth}
\newlength{\dinmargin}
\title{Localization in Lattice Gauge Theory and a New Multigrid Method
}
\author{Martin B\"aker \protect\\
Universit\"at Hamburg, II.~Inst.~f.~Theor.~Physik\protect\\
Luruper Chaussee 149 \protect\\
D-22761 Hamburg \protect\\
\small e-mail : $<$i02bae@dsyibm.desy.de$>$
                      }
\newcommand{\stress}{\sl}
\newcommand{\matr}[1]{{\bf #1}}
\newcommand{\op}{{\bf L}}
\newcommand{\vect}[1]{{\bf #1}}
\newcommand{\calA}{{\cal A}}

\newcommand{\calH}{{\cal H}}
\newcommand{\calR}{{\cal R}}
\newcommand{\kernel}{{\calA^{[ 0\,j]}}}
\newcommand{\kernelmac}[1]{{\calA^{[ 0\, #1]}}}
\newcommand{\kernelk}{\kernelmac{k}}
\newcommand{\kernelone}{\kernelmac{1}}
\newcommand{\kernelarg}{{\calA^{[ 0\,j]}_{zx}}}
\newcommand{\kernelvec}{{\calA^{[ 0\,j]}_{\cdot,x}}}
\newcommand{\kernelmacarg}[1]{{\calA^{[ 0\,#1]}_{zx}}}
\newcommand{\kernelmacvec}[1]{{\calA^{[ 0\,#1]}_{\cdot,x}}}
\newcommand{\kernelmacvar}[2]{{\calA^{[ 0\,#1]}_{#2}}}
\newcommand{\mittel}{{C^{[ j\,0]}}}
\newcommand{\mittelk}{{C^{[ k\,0]}}}

\newcommand{\resid}{{\calR^{[ 0\,j]}_{\cdot,x}}}
\newcommand{\SUZWEI}{{\rm SU(2)}}
\newcommand{\cuberes}{|_{[x]}}
\newcommand{\Laplace}{\bigtriangleup}

\newcommand{\LGt}{Lattice Gauge Theory}
 
\def\tilde#1{\widetilde{#1}}
\def\xitilde{{\tilde{\vect{\xi}}}}
\def\rho{\varrho}
\def\epsilon{\varepsilon}
\def\phi{\varphi}
\def\be{\begin{equation}}
\def\ee{\end{equation}}
\begin{document}
\begin{titlepage}
\begin{center}
\vspace*{5cm}
{ \LARGE Localization in Lattice Gauge Theory and a New Multigrid Method
\\}
\vspace{1.5cm}
{\LARGE Martin B\"aker \protect\\
\vspace{1cm}
\Large Universit\"at Hamburg, II.~Inst.~f.~Theor.~Physik\protect\\[6pt]
Luruper Chaussee 149 \protect\\[8pt]
D-22761 Hamburg \protect\\[8pt]
\small e-mail : $<$i02bae@dsyibm.desy.de$>$
\\                      }
\end{center}
\vspace{3cm}
\begin{abstract}
We show numerically that the lowest eigenmodes of the 2-dimensional
Laplace-operator with \SUZWEI\ gauge couplings are strongly localized.
A connection is drawn to the Anderson-Localization problem.
A new Multigrid algorithm, capable to deal with these modes,
shows no critical slowing down for this problem.
\end{abstract}
\end{titlepage}
%
%
It is well-known that the convergence of local relaxation algorithms
for solving inhomogeneous linear equations of the form
$\matr{L} \xi = \vect{f}$
is determined by the lowest eigenmodes of the problem matrix.

We studied these modes in an example coming from \LGt, using
\be \matr{L} = -\Laplace + \delta m^2 - \epsilon_0 \quad.
                                      \label{covLap}     \ee
Here $\Laplace$ is the 2-dimensional covariant Laplace operator,
i.e.~the discretized Laplace operator with \SUZWEI-matrices
as nearest-neighbour-couplings (see sect.~\ref{ModelProblem} for
a more thorough explanation of this), $\epsilon_0$ is its lowest
eigenvalue and $\delta m^2$ is the critical parameter. The lowest
eigenvalue has
to be subtracted to make the problem critical---otherwise there
would be no critical slowing down and no need to apply a multigrid.
This subtraction of a constant does obviously
not affect the shape of the eigenvectors.
Physically, our model corresponds to a Higgs-doublet with a
\SUZWEI-charge.

It is found
that the lowest eigenmodes of this model
are strongly localized, i.e.~they are appreciably large
only in a small region of the grid. This result will be described in
the first part of this paper. In the second part a
recently proposed multigrid algorithm \cite{previous}
and its
performance for the model problem is investigated. This algorithm works
extremely well (it eliminates critical slowing down completely),
because it is able to handle the localized modes.

\setcounter{equation}{0}
\section{Localization}
\subsection{Localized modes in a model problem}
\label{ModelProblem}

The fundamental form of an inhomogeneous equation is
\be \matr{L} \xi = \vect{f}\quad.\ee
If we choose $\matr{L}$ as in eq.~(0.1), our equation
becomes the propagator equation for a bosonic particle in a
\SUZWEI-gauge field background. The covariant Laplace-operator in
stencil-notation is \be \Laplace(z) = \left[\matrix{0&U_{z,2}&0\cr
U_{z,-1}&4&U_{z,1}\cr 0&U_{z,-2}&0\cr} \right]\quad,\label{stencil}\ee
with $U_{z,\mu} \in \SUZWEI$. The second index denotes the direction of
the coupling to the neighbour.
The link matrices $U_{z,\mu}$ fulfill $U_{z,-\mu} = U_{z-\mu,\mu}^*$.
They are distributed according to the Wilson action
\cite{Wilson}
\be S_W = \frac{\beta}{4}
                \sum_{P} {\rm Re\ tr\ }\left(1-U_P \right)
\quad.\ee Here
 $\beta = 4/g^2$ is the inverse coupling and the sum is over all
Plaquettes in the lattice. $U_P$ denotes the parallel transport
around the Plaquette
\begin{center} 
\begin{picture}(300,50)(0,0)
%
%
\thicklines
\put(150,30){\mbox{$U_P = U_{z,\nu}^*
                  U_{z+\nu,\mu}^* U_{z+\mu,\nu} U_{z,\mu}$}}
\put(150,10){\mbox{$\mu \ne \nu$}}
%
%
\multiput(20, 10)(25,0){2}{\circle*{4}}
\multiput(20, 35)(25,0){2}{\circle*{4}}
%
%
\put(20,10){\vector(1,0){23}}
\put(45,35){\vector(-1,0){23}}
\put(45,10){\vector(0,1){23}}
\put(20,35){\vector(0,-1){23}}
%
%
\small
\put(25,0){\mbox{$U_{z,\mu}$}}
\put(0,20){\mbox{$U_{z,\nu}^*$}}
\put(50,20){\mbox{$U_{z+\mu,\nu}$}}
\put(25,43){\mbox{$U_{z+\nu,\mu}^*$}}
%
%
\put(90,10){\vector(0,1){20}}
\put(90,10){\vector(1,0){20}}
\put(112,7){\mbox{$\mu$}}
\put(95,28){\mbox{$\nu$}}
\end{picture} \end{center}

This distribution leads to a correlation between
the gauge field matrices with finite correlation length $\chi$
for finite $\beta$. The case
$\beta = 0$ corresponds to a completely random coice of the matrices
($\chi=0$),
for $\beta=\infty$ all matrices are $\matr{1}$
($\chi=\infty$). In this sense, $\beta$
is a disorder parameter, the smaller $\beta$ the shorter the correlation
length and the larger the disorder.

Now we want to study the lowest---and the highest---eigenmodes
of this operator. If we look only at the norm of the eigenmodes
we can see immediately that the lowest and the highest eigenmodes
will look identical because of the following

\begin{quote}
{\bf Theorem: \/} Let $\bf L$ be a $(n\times n)$-matrix with the
following properties:
\begin{itemize}
\item $\bf L$ has a constant diagonal $\alpha$,
\item For all $i,j$ with $i\ne j$ and $i+j$ even: $L_{ij}=0$.
\end{itemize}
Then the following statement holds:

If $\xi^k$ is an eigenvector of $\bf L$ to the eigenvalue $\lambda^k$
($k<n/2$),
then the vector $ (-)^j \xi^k_j$ is also an eigenvector to
$\lambda^{n-k} = 2 \alpha - \lambda^k$.
\end{quote}

This theorem is also true if the matrix elements are matrices
themselves. It applies to the Laplace operator if its matrix
elements are ordered in the right way (checkerbord fashion).

Fig.~2 shows the norm of the
                 lowest eigenmode of the covariant Laplace operator
on a $64^2$-lattice at $\beta=1.0$  and
on a $32^2$-lattice at $\beta=5.0$ .
For the smaller $\beta$-value the localization can be seen clearly,
for larger $\beta$ the localization is in a more extended domain.

This of course poses the question whether all modes are localized.
To answer this, we calculated all modes on a smaller grid, using a
standard library routine. It was found that only a few of the low modes
show localization. It might be possible that on very large grids (with
very low values of~$\beta$) many of the modes are localized, as
localization increases with the disorder. But to study this a large
computational effort would be needed. (The simple storage of all
eigenvectors on a $128^2$-lattice would need about 1 Gigabyte.)

It is expected that the sharpness of localization will increase with
the disorder, i.e.~with decreasing $\beta$. To study
this effect we may look at the {\stress participation ratio}
defined as \cite{Edwards}
\be \alpha^{-1} = {1\over N}
                   \frac{\left(\sum_{i=1}^N |\xi_i|^2\right)^2}
                   {\sum_{i=1}^N |\xi_i|^4}\quad,\ee
where $N$ denotes the number of grid points.
This quantity measures the fraction of the lattice over which
the peak of the localized state~$\xi$ is spread. Fig.~3 shows
the participation ratio of the lowest eigenmode
as function of $\beta$ calculated on
$64^2$-lattices. The dependence of
the disorder can be seen clearly. The
absolute size of the localized state, measured by $N \alpha^{-1}$,
does not depend on the grid size if the grid is larger than the
localization peak.

We want to remark here that a similar phenomenon of localization
also has been found for the two-dimensional Dirac-equation in a
\SUZWEI-gauge-field \cite{baeker2}
and for the same two models in four dimensions \cite{ThomasLattice}.

\subsection{Explaining the localization}
How can we understand this phenomenon of localization in
non-abelian gauge fields?

We will try to relate it to
Anderson-localization \cite{And,Thou2,Nago}.
Anderson examined a tight-binding-model of an
electron in a random potential $V(z)$, leading to a Schr\"odinger
equation
\be (-\Laplace_{\rm scalar} + V(z)) \psi = E \psi \quad.\ee
Localization
of all modes may occur, dependend on the disorder and the dimension.
In one dimension, arbitrary small disorder leads to localization,
in two dimensions there is a phase transition between weak and strong
localization as one increases the disorder, in higher dimensions a
transition between non-localized and localized states occurs.
A theoretical understanding of this was achieved in the papers
\cite{Froeh}. As the Schr\"odinger operator
for this model is again the Laplacian (without a gauge field), we may
                                     expect a similarity
between our localization problem and Anderson localization,
but there are crucial differences: First, our operator is not fully
random, because the equilibrated gauge field possesses correlations.
Second, our operator shows {\stress off-diagonal disorder\/}: It is
not the potential that varies from site to site, but the couplings
between the sites. And third, the couplings do not vary in {\stress
strength,\/}
but in orientation in colour space, resulting in a frustrated
system.

Nevertheless, in the following we will try to stress similarities
between the two models, giving us an---at least intuitive---picture
of what happens.

We have to look for a quantity in our model that may play the role of
the random potential $V(z)$ in the Anderson problem. There we can
extract the value of the potential by calculating the difference
between the diagonal element and the norm of the couplings, because
this is zero for the Laplace operator without potential, see
eq.~(\ref{stencil}).
Doing this for our model problem of course only gives
$\delta m^2-\epsilon_0$,
independend of the lattice site. To arrive at a varying quantity,
we may look at a blocked version of our operator.
The simplest blocking procedure one can think of is the ``taxi-driver''
blocking \cite{Balaban}.
Here four points are blocked to one and all block-quantities
are calculated by parallel-transporting the fine-grid-quantities
to one of the four points (e.g.~the lower left point)
via the shortest possible route. (For the upper right point there are
two routes, each weighted with a factor $1/2$.)

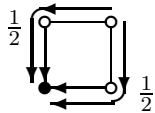
\begin{center} 
\begin{picture}(100,50)(0,0)
%
%
\thicklines
%
%
\multiput(20, 10)(25,0){2}{\circle{4}}
\multiput(20, 35)(25,0){2}{\circle{4}}
\put(20,10){\circle*{4}}
%
%
\thinlines
\put(43,35){\line(-1,0){21}}
\put(45,33){\line(0,-1){21}}
\thicklines
\put(43,10){\vector(-1,0){21}}
\put(20,33){\vector(0,-1){21}}
\put(45,10){\oval(10,10)[br]}
\put(20,35){\oval(10,10)[tl]}
\put(45,40){\vector(-1,0){27}}
\put(50,35){\vector(0,-1){27}}
\put(15,36){\vector(0,-1){24}}
\put(46, 4){\vector(-1,0){24}}
\put(5,30){\mbox{$\frac12$}}
\put(55,5){\mbox{$\frac12$}}
\end{picture} \end{center}
If we use a blocking-operator of this type and calculate
$\op_{\rm block}=\mittel \cdot(-\Laplace)\cdot \kernel$,
the blocked operator, we still have
an operator with fluctuating bonds and a constant diagonal. But now
the bonds are fluctuating in strength, so we have to separate the
kinetic and the potential part of the diagonal elements by
calculating
   $ V(x) = \op_{\rm block}(x,x)-\sum_y \|\op_{\rm block}(x,y)\|$
as explained above.
This quantity now {\stress is} fluctuating on the block lattice, and so
we arrived at a situation much more similar to Anderson-localization.

The blocking procedure has a certain arbitrariness (e.g.~in
the choice of the blocks), but a simple calculation shows that it is
mainly the quantity $\|F_{\mu\nu}(z)\|$ that enters into $V(x)$.
(This has to be expected taking into
account that the quantity involved has to be gauge invariant.) Let us
now look at the
field strength norm, or, which is more conclusive, at the quantity
$W(z) = \sum_{\mu \ne \nu} F_{\mu \nu}(z) F^{\mu \nu}(z)$.
Fig.~4 shows this quantity for the same configuration as in
figure~2, bottom,
and one can see clearly that the localization is in a region where
$W(z)$ is small, as should be expected from our argument. This is
true for {\stress all\/} configurations we looked at:
{\stress The localization center is always in a region with low field
strength sum.\/}

We can support this idea further by looking at the eigenmodes of
the following operator, which we called Anderson-Laplace-operator:
$D_{AL} = - \Laplace_{\rm scalar} + W(z)$,
where $\Laplace_{\rm scalar}$ is the
Laplace operator without gauge field.
So now we are looking at a true Anderson-localization
problem, except that the random potential is not independend at
different sites. There is a finite correlation length, as explained in
the previous section.
Fig.~5 shows the lowest mode of this operator.
If compared to fig.~1, bottom,
one sees that the center of localization sits at the same place.

In conclusion we can say that our analysis shows that firstly
the lowest eigenmodes of the covariant Laplace operator in a
\SUZWEI-gauge field are localized and that secondly this
localization occurs where the field strength norm is small.
This quantity can be interpreted as a random potential on a block
lattice. In this way we were able to draw a connection to
Anderson localization.
%
%
\setcounter{equation}{0}
\section{The Iteratively Smoothing Unigrid}
\subsection{The problem}
Disordered systems are among the most interesting and most difficult
models in physics. Here we are interested in the numerical
solution of---discretized---differential equations for such models.
(An example for this is the
inversion of the fermion matrix.) As the critical point of the system is
approached, simple local algorithms face the problem of {\stress
critical slowing down}, that is, the nearer one gets to the critical
point the slower the convergence.

The use of nonlocal methods may overcome this problem. For the solution
of ``ordered'' problems multigrid methods have been very successfull.
Even in the disordered case the ``Algebraic Multigrid'' (AMG)
\cite{Ruge} was
applied with great success to scalar problems like the ``random resistor
network'' \cite{Edw}.
But up to now, no generalization of this method has been
found that could be used for \LGt.

In a previous paper \cite{previous}
                    this problem was tackled and a new algorithm was
proposed. Here an improved version will be explained. It will be applied
to the model problem described above. This algorithm
is a {\stress unigrid\/} algorithm. To prepare for the following
consideration, in the next section the multigrid method will be
reformulated in the unigrid language.

\subsection{The Unigrid}
Suppose the equation to solve lives on a ``fundamental'' lattice
$\Lambda^0$ with lattice constant $a_0$. We write the equation as
\be \op_0 \xi^0 = f^0 \quad ,\label{fundeq2}\ee
                               where the index 0 tells us that it is
formulated on the fundamental lattice. Inhomogeneous equations
of this type also arise when an eigenvalue equation is solved
by inverse iteration \cite{Press}. We will use inverse
iteration heavily in our algorithm later on.

One now introduces auxiliary lattices $\Lambda^1,\Lambda^2,\ldots
\Lambda^N$, called layers,
           with lattice constants $a_j = L_b^j a_0$, where $L_b$ is
the blocking factor (typically, one chooses $L_b=2$).
The last lattice $\Lambda^N$ consists of only one point.
The different ways of transferring information between the lattices
makes the crucial distinction
between a unigrid and a true multigrid method.
Let $\calH^j$ be the space of functions on lattice $\Lambda^j$. Then
we introduce {\stress grid transfer operators:}
\be {\rm the\ interpolation\ operator:\quad }
                                     \kernel : \calH^j \mapsto \calH^0
{\quad \rm and}\ee
\be {\rm the\ restriction\ operator:\quad }
                                   \mittel : \calH^0 \mapsto \calH^j
{\quad .}\ee
In a true multigrid grid transfer operators are composed from
transfer operators that act between adjacent layers. From this it
is quite clear
that every multigrid can be formulated as a unigrid (instead of
transferring information directly
from $\Lambda^j$ to $\Lambda^k$ go first to the fundamental lattice
and afterwards to the target space), but not vice versa.
{}From the point of view of computational complexity, a unigrid method
is inferior because an equal amount of work needs to be done on
all layers.

\subsection{The meaning of smoothness}
The basic principle of multigrid algorithms (for ordered elliptical
problems) originates from the observation that
after doing local relaxation sweeps, the error gets {\stress smooth}.
Hence it should be possible to represent the error on a coarser
lattice, because the intermediate values can be obtained by smooth
interpolation. ``Good'' interpolation operators are therefore known
beforehand; for example, one may use linear interpolation.
Normally, the smooth modes are the low-lying
eigenmodes of the operator.

How can we extend this observation to disordered problems, where
no a priori notion of smoothness is given?
As explained in \cite{previous}, we propose the following definition
of smoothness in a disordered system. It assumes that a fundamental
differential operator $\op_0$ specifies the problem. In this case

\begin{quote}
{\bf Definition: } A function $\xi$ on $\Lambda^0$ is {\stress smooth on
length scale $a$} when
\be \|\op_0 \xi\|^2 \ll \|\xi\|^2  \quad ,\ee
in units $a=1$.
\end{quote}

This definition implies that the smoothest function is the lowest
eigenmode of $\op_0$.\footnote{It has recently been remarked by Sokal
\cite{Sokal}
that the operator $\op$ does not possess eigenvectors in a strict
sense, because it maps a space on its dual and there does not exist
a natural scalar product on the two spaces. He proposes to look instead
at $B_0^{-1} \op$, where $B_0$ is defined by the relaxation algorithm.
In case of a matrix with constant diagonal this will not make any
difference (as long as we use e.g.~Jacobi-iteration), so this does not
give problems with our model problem. Remark that this is not true for
a true multigrid, because there on a block lattice a non-constant
effective mass might be generated. In such an algorithm, the definition
of smoothness should take the relaxation algorithm into account,
as it is done in the AMG-context.}

So we arrive at the basic principle stated above
even for the disordered case: The slow-converging modes,
which have to be represented on coarser grids, are smooth. Of course
we still have to show that the definition is sensible and can be used
to construct a good algorithm.

As a first step we can see that the notion of {\stress algebraic
smoothness} as introduced in the context of the AMG implies smoothness
in our sense. The error $\vect{e}$ of an approximative solution
of the differential equation is called ``algebraically smooth'' if
the residual $\vect{r}$ is small compared to it, so
\be \vect{r}_i=\sum_j \op_{0,ij} \vect{e}_j \ll \vect{e}_i
\Longrightarrow \| \op_0\,\vect{e} \|^2 \ll \| \vect{e} \|^2 \quad .\ee

The crucial step in the setup of the algorithm is the choice of the
grid transfer operators $\kernel$ and $\mittel$. These operators
should be smooth in our sense, because we want to use them to represent
a smooth error on a coarser grid. But because our definition of
smoothness depends on the problem matrix $\op$, they are not given
a priori. Instead, we have to {\stress compute} these operators.

In the following, we will adopt the
{\stress Galerkin choice\/} $\mittel = \kernel^*$, where ${}^*$ denotes
the adjoint, and the {\stress coarse-grid-operator\/} $\op_j$ will
be defined by $\op_j = \mittel \,\op_0 \,\kernel$.
                                            So we only need to construct
good interpolation operators.

\begin{figure}
\begin{center} \setlength{\unitlength}{1.35pt}
\begin{picture}(350,100)(0,0)
%
%
\thicklines
%
%
\multiput(150, 10)(10,0){9}{\circle*{3}}
\multiput(150, 20)(10,0){9}{\circle*{3}}
\multiput(150, 30)(10,0){9}{\circle*{3}}
\multiput(150, 40)(10,0){9}{\circle*{3}}
\multiput(150, 50)(10,0){9}{\circle*{3}}
\multiput(150, 60)(10,0){9}{\circle*{3}}
\multiput(150, 70)(10,0){9}{\circle*{3}}
\multiput(150, 80)(10,0){9}{\circle*{3}}
\multiput(150, 90)(10,0){9}{\circle*{3}}
%
%
\thinlines
\multiput(156,16)(40,0){2}{\dashbox{3}(28,28){}}
\multiput(156,56)(40,0){2}{\dashbox{3}(28,28){}}
\multiput(154,34)(40,0){2}{\dashbox{2}(32,32){}}
\put(176,36){\dashbox{3}(28,28){}}
\put(174,14){\dashbox{2}(32,32){}}
\put(174,54){\dashbox{2}(32,32){}}
\put(152,12){\dashbox{5}(76,76){}}
\end{picture} \end{center}
\caption{Supports of the interpolation operator for the layers~1
and 2. On layer~1, more than one support is drawn to show
the overlap.   \label{support}}
\end{figure}
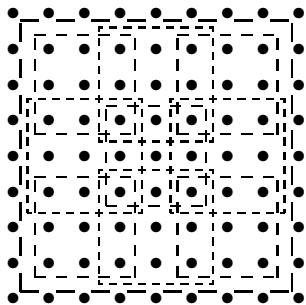
We have to restrict the operators to a part of the lattice, because
otherwise interpolation would be too costly.
For the time being we will choose the supports of the interpolation
operators to be fixed blocks $[x]$ as shown in figure~\ref{support}.
Therefore the
operators will possess representations as rectangular matrices
with elements $\kernelarg$ with $x \in \Lambda^j$ and $z \in \Lambda^0$.
The matrix $\kernel$ contains all interpolation operators on layer~$j$.
The single interpolation operator on block $[x]$ is denoted by
$\kernelvec$. It corresponds to one column vector of the matrix
operator $\kernel$.
$\kernelvec$ vanishes outside the block $[x]$.
It has been found that there is little hope
in eliminating critical slowing down for the inversion of the
Dirac-operator without overlapping blocks \cite{ThomasDiss}.
As we intend to apply our algorithm also to this case, we choose
overlapping blocks as shown in the figure.

We now want to look for smooth interpolation operators in the specified
sense which fulfill (approximately) the eigenvalue equation
restricted to the block $[x]$
\be  \op_0|_{[x]} \,\kernelvec =
                     \epsilon_0(x)\, \kernelvec \quad .\label{eveq}\ee
Here $\op_0|_{[x]}$
        denotes the restriction of $\op_0$ to the block $[x]$.
As the interpolation operator must vanish outside the block, we impose
Dirichlet boundary conditions.
The crucial assumption of our algorithm is that the solution
$\kernelvec$ of this equation is smooth on length scale~$a_j$.
This is true for the scalar Laplace-operator, where the solution is
half a sine wave on the block.

If we know such interpolation operators
we can start a unigrid algorithm in the usual way: After relaxing
on the fundamental layer the error~$\vect{e^0}$ of the approximative
solution $\xitilde^0$ defined as $\vect{e^0} = \xi^0 -\xitilde^0$
should be smooth. It fulfills the error equation
                                     $\op_0 \vect{e}^0=\vect{r}^0$,
where $\vect{r}^0 = \vect{f}^0 - \op_0 \xitilde^0$ is the
{\stress residual}.

As the error is smooth, it can be obtained by smooth interpolation of
a function $\vect{e}^1$ living on $\Lambda^1$:
\be \vect{e}^0 = \kernelone \,\vect{e}^1 \quad.\ee
Inserting this into the error equation yields
\begin{eqnarray}
    \op_0 \, \kernelone \, \vect{e}^1& =& \vect{r}^0\\
 C^{[1\,0]}\,
    \op_0 \,\kernelone\, \vect{e}^1& =&C^{[1\,0]}\, \vect{r}^0 \\
    \op_1 \,\vect{e}^1& =& \vect{r}^1 \label{blockeq}\quad,
\end{eqnarray}
which involves only functions and operators on the block lattice.
After relaxing on eq.~(\ref{blockeq})
                         we interpolate our estimate $\vect{e}^1$ back
to the fundamental lattice and replace our approximation by a
better one: $\xitilde \leftarrow \xitilde + \kernelone\, \vect{e}^1$.
Now the error is expected to
                   be smooth on the length scale $a_1$ because we have
relaxed it on $\Lambda^1$, and so we can proceed to the next layer.
When we come to layer $\Lambda^j$, the error should be smooth on
scale $a_{j-1}$. After relaxing on this layer, we correct our
approximation again:
$\xitilde \leftarrow \xitilde + \kernel\, \vect{e}^j$.
By this we have smoothened the error on the larger scale $a_j$.
Eventually we will reach the lattice~$\Lambda^N$ where we can solve
the equation exactly and can thereby remove the smoothest mode from the
error.
The difference to a true multigrid can be seen clearly: The
multigrid is defined in a recursive way. (\ref{blockeq})
                                                is {\stress solved}
by going to the next-coarser lattice without correcting the
approximate solution~$\xitilde^0$ on the fundamental lattice.

\subsection{The ISU-algorithm}
But of course the question is: {\stress ``Where do we get the smooth
operators?''\/} Consider for instance
$\kernelmac{N}$. It should satisfy the equation
\be \op_0\, \kernelmacvar{N}{\cdot} = \epsilon_0\,
\kernelmacvar{N}{\cdot} \quad, \label{eveqN}  \ee
involving the full unrestricted operator~$\op_0$, because the
last lattice consists of only one point. If we want to solve this
equation with inverse iteration (i.e.~by computing
$\op_0^{-n} \kernelmacvar{N}{\cdot,\,\rm start}$
                                         for large $n$ and an arbitrary
starting vector), we will have to solve an equation which seems to
be exactly as difficult as our starting point, eq.~(\ref{fundeq2}).

But this is not so. The worst-converging mode of our starting
equation is the mode to the lowest eigenvalue of $\op_0$, but now
we want to {\stress compute\/} this mode, so it does not contribute
to the error of the eigenvalue equation. (We have to do a
simple normalization step after each iteration.)
Consequently the mode to the second-lowest
eigenvalue of $\op_0$ is the one that converges worst and if we could
handle this (and all higher modes
as well), we could also handle the lowest mode in our
inhomogeneous equation~(\ref{fundeq2})
                          by solving first eq.~(\ref{eveqN}).

The basic idea of our algorithm is that the higher modes are smooth
on shorter length-scales. This means that it should be possible to
construct them out of pieces which are smooth on these length scales
and have supports on parts of the lattice. So the next-lowest modes
$\xi^{\rm low}$
are representable by linear combination of the interpolation operators
$\kernelmacvec{N-1}$:
\be \xi^{\rm low}_z =
  \sum_{x\in \Lambda^{N-1}} c_x \kernelmacarg{N-1} \quad.    \ee
If this is true---and it is for our model
problem---we see that the calculation of $\kernelmac{N-1}$
is similar. Again the worst-converging mode is the mode we aim at,
the next-higher modes can be represented on smaller blocks.
Their calculation is therefore simpler. Finally we arrive at the
calculation of $\kernelmacvec{1}$, having to solve an equation on
a~$3\times 3$-lattice. This is easily done. Because of the
Dirichlet boundary conditions there is no low-lying mode
here.

{\bf Remark:\/}
It might happen that the lowest eigenvalue is much larger than
the difference between it and the next eigenvalue. In this case, many
inverse iterations have to be done to resolve the two corresponding
modes. A possible remedy for this problem is to calculate estimates
$\lambda$ of the lowest eigenvalue as we proceed and to invert not
$\op_0$ but $\op_0 - \lambda $. This problem will not arise on
the last layer, because otherwise the operator would not be critical
and simple relaxation algorithms will suffice to solve the equation.

We identify the site $x\in \Lambda^j$ with the block $[x]$ in
$\lambda^0$ having $x$ at the center. Eq.~(\ref{eveq}) is an
eigenvalue equation on~$[x]$ for the vector $\kernelvec$.
                              It can be solved via inverse iteration
by our unigrid method, using the already calculated interpolation
operators $\kernelmacvar{k}{\cdot,y}$ with supports inside the block.
With this we arrive at the following

\bigskip
{\bf Algorithm for calculating smooth interpolation operators:}
\smallskip

{\bf For} $1\le j \le N$ {\bf do}

{\bf \ For} all $x \in \Lambda^j$ {\bf do}
\begin{enumerate}
\item Choose initial value $\kernelmacvar{j}{ \cdot,x;\,{\rm start}}$.
\item Relax on the fundamental lattice on block $[x]$.
\item {\bf For} all $1\le k <  j$ {\bf do}:\\
Calculate the residual $\resid=\op_0\cuberes\,\kernelvec -
\kernelmacvar{j}{\cdot,x;\,{\rm start}} $ .\\
Block the residual to layer $\Lambda^k$: $\resid^\prime
  = \kernelmac{k} \,\resid$.\\
Calculate $\op_k\cuberes=
                \mittelk\,\op \cuberes\,
            \kernelk$.\\
Determine approximate solution of
$\op_k\cuberes \,\delta\kernelvec^\prime
 = \resid^\prime$ by relaxation on $[x]\cap \Lambda^k$.\\
Correct $\kernelvec: \quad
\kernelarg \longleftarrow \kernelarg +
  \sum_{y\in \Lambda^k}    \kernelmacvar{k}{z,y}\,\delta
\kernelmacvar{j}{y,x}^\prime \quad $.
\item Normalize the interpolation operator.
\item {\bf If} approximate solution $\kernelvec$ good solution of
               the eigenvalue equation {\bf then do next\/} $x$\\
      {\bf If} approximate solution $\kernelvec$ good solution of
               the inhomogeneous eq.

      {\bf \ then} set $\kernelmacvar{j}{\cdot,x;\,{\rm start}}
                                        \leftarrow \kernelvec$
\item {\bf go to } 2.
\end{enumerate}
We call this method {\stress
Iteratively Smoothing Unigrid\/} or ISU, because it is a unigrid
method which computes smooth operators by means of an iterative method
(and not directly from the given operator as in the AMG-algorithm).

Now we could try to use the same method to define a true multigrid
algorithm: Just replace the eigenvalue equation for $\kernelvec$ by
an equation for $\calA^j_{\cdot,x}$
                           which interpolates between adjacent layers
and use $\op_{j-1}$ as the operator for this equation. In this way,
we would get operators which are smooth with respect to the blocked
differential operator. But this will often
                                     not work. To see this, formulate
the new true multigrid algorithm in unigrid language. It involves
operators $\kernel = \calA^1 \cdot \calA^2 \cdots \calA^{j}$. But
the product of operators which are smooth on different length scales
is smooth only on the shorter of these length scales.
So we will never get a transfer operator that is smooth on large
scales. To put the same fact in another way: In a true multigrid
as used for ordered problems, interpolation operators
$\calA^j$ will smoothly interpolate functions that are smooth on
{\stress all}
length scales $a_k > a_j$ (the usual linear operators for example
interpolate constants to constants, regardless of the scale), but
in our case $\kernel$ will not be able to do this.

This tells us that our algorithm really is a unigrid algorithm, not
just a multigrid in disguise. It is therefore impossible to apply the
usual two-grid-analysis to prove convergence. Furthermore, we can not
stop the algorithm on a layer $j<N$, because in this case the
modes on the larger scales would not be handled appropriately.

{}From the above description it is clear that the work involved in
calculating good interpolation operators is larger than the time
needed for the solution of the equation~(\ref{fundeq2}) itself.
The following table shows the
computational costs of the algorithm, compared to a true multigrid
and to a local relaxation. Here, $L$ denotes the grid length and
$d$ is the dimension.

\begin{center}
\begin{tabular}{|l|l|l|}\hline
  Algorithm           & CPU-time    &Storage space\\ \hline
 Local  relaxation      & $ L^{d+z} (z\approx 2)$    & $L^d$ \\
 True multigrid and AMG& $ L^d $           & $L^d$         \\
ISU-algorithm& $L^d\ln^2 L$        & $L^d \ln L$ \\ \hline
\end{tabular}
\end{center}
\subsection{Performance of ISU}
We studied this algorithm for the 2-dimensional bosonic model
problem, as described in eqs.~(0.1) and (2.1).
                                The subtraction of the lowest eigenvalue
makes the problem critical, and we can directly control criticality
by tuning $\delta m^2$, the lowest eigenvalue of the full problem.

We measured the inverse asymptotic convergence rate~$\tau$ defined
as
\be \tau = \lim_{n\rightarrow \infty} \frac{-1}{\ln \rho_n} \quad,\ee
where $\rho_n$ is the quotient of the error norms before and after
iteration number~$n$. For large~$n$, this quantity approaches a
constant.

Fig.~6 shows the inverse
asymptotic convergence rate as
a function of~$\delta m^2$ for grid sizes $32^2$--$128^2$ at $\beta=1.0$
in a \SUZWEI-gauge field background. Absence of critical slowing down
can be seen clearly, and the absolute value of~$\tau$
is quite small. ($\tau=1$ corresponds to a reduction of the residual
by a factor of $e$ in one multigrid sweep.) The sweeps are V-cycles
with one pre- and one post-relaxation step.
The results do not vary appreciably by changing~$\beta$.

{\bf Remark:\/} To conclude that critical slowing down is absent,
it is not sufficient to study only the dependence of $\delta m^2$
for fixed grid sizes. Our method ensures correct treatment of
the {\stress lowest} mode with eigenvalue $\delta m^2$.
But the eigenvalue
of the second-lowest mode depends on the grid-size, so the grid-size
has to be large enough to make also this eigenvalue fairly low.

Because of the large work involved in calculating the interpolation
operators, it has also to be
veryfied that the number of inverse iterations needed for the
calculation of the interpolation operators does not grow with the
grid size. We found that six multigrid sweeps with one pre- and no
postrelaxation step were sufficient for this on every layer,
{\stress regardless of the lattice size}; doing more sweeps and
thereby calculating the operators more exactly did not improve
the convergence of eq.~(\ref{fundeq2}).

So it is clear that there is no critical slowing down for
the solution of the model problem. Nevertheless the question
has to be answered at which grid sizes this will pay off, since the
overhead is large. A careful comparison with the conjugate gradient
algorithm on CPU-time grounds will have to be done. We think this
will be worthwile only if the ISU works as well for fermions.

To understand this success we can investigate the algorithm more
closely. To this end we have calculated {\stress all} eigenmodes
of the covariant Laplace operator~(0.1)---using a standard
                                 library function---and
a solution of eq.~(\ref{fundeq2}). Now we were able to start the
algorithm with an initial value of which we knew the error in advance,
and to monitor the behaviour of the error as the algorithm proceeds,
always expanding the error into the eigenmodes. So we could check
that the fundamental relaxation indeed smoothens the error by
eliminating the contribution of the higher modes. The coarser the
lattice becomes the lower are the modes which are eliminated, until
on the last lattice the contribution of the lowest mode is set
exactly to zero.

We also checked that it is indeed possible to represent the low-lying
modes by the interpolation operators. We
                       calculated that part of the modes which
was orthogonal on all interpolation operators
                      on a given layer. This quantity
was small, so the overlap between low-lying modes and interpolation
operators is large.

This latter result is not surprising, as we already know that the
lowest modes are localized. If a mode consists of a few localized
parts it seems clear that we can patch it together from operators
which are restricted to a part of the grid. This suggests that
localization of low-lying modes is a good prerequisite for
convergence of our algorithm.

Fig.~7 shows the interpolation operators $\kernelmac{N-1}$
and $\kernelmac{N-2}$
belonging to the same gauge field configuration as Fig.~2,
bottom. It can be
clearly seen that the different support boundaries are able to single
out the different localization centers. On smaller blocks, the
operators are centered in the middle of the block, because here the
restriction through the boundary is too strong.

It is crucial for this possibility of representing the localized modes
by the interpolation operators that there exists a block into which
the modes fit well. This is not true
for other models of disordered systems. Our algorithm was also tried
on the ``random resistor problem'' \cite{Thou1},
                                    but here the simple method
of fixing the blocks geometrically led to critical slowing down
($z\approx 0.7$). This is explained by the fact that
there existed parts of eigenmodes which did not fit into any
of the blocks and could therefore not be combined from the interpolation
operators. (We studied this by again calculating all modes and
looking at the shape of the bad-converging ones.) Nevertheless
it is possible to generalize our algorithm by choosing the blockcenters
in a more sophisticated way, using an algorithm of AMG-type for this
step, and then applying the ISU-algorithm. This will be done in the
future, using a C++-class library which is developed at the moment
\cite{Speh}.

\section{Conclusions}
We have seen the localization of the lowest (and highest)
modes in a 2-dimensional \LGt. This phenomenon could be understood
by connecting it to the problem of Anderson-Localization.

The performance of a new Multigrid algorithm was studied. For the model
problem it showed {\stress no\/} critical slowing down. The algorithm
is well suited to deal with low-lying localized modes, because good
approximations to these modes are used for the interpolation.

The next step will be to investigate localization and the behaviour of
the ISU-algorithm for the Dirac-equation. This work is in
progress.
\section*{Acknowledgements}
I wish to thank Gerhard Mack for his constant support and many
inspiring discussions. Further thanks are due to S. Meyer and M. Speh
for their help with running the ISU on different machines, and to
A. Brandt, S. Solomon and all members of the Hamburg Multigrid Group
for helpfull discussions.

Financial support by the Deutsche Forschungsgemeinschaft ist gratefully
acknowledged.
%
%

\clearpage
\end{document}